\documentclass[%
 reprint,
superscriptaddress,
 amsmath,amssymb,
 aps,
 prl,
]{revtex4-1}

\usepackage{graphicx}
\usepackage{dcolumn}
\usepackage{bm}
\usepackage{natbib}
\usepackage{hyperref}
\usepackage{siunitx}
\usepackage[caption=false]{subfig}
\usepackage{booktabs}
\usepackage[export]{adjustbox}
\usepackage{multirow}
\usepackage{physics}
\usepackage{accents}
\usepackage{color}
\usepackage{xcolor}

\usepackage{multibib}
\usepackage[]{xcolor}

\newcommand{\SectionPRL}[1]{\emph{#1.---~}}%

\begin{document}


\title{Fundamental Limits of Feedback Cooling Ultracold Atomic Gases}

\author{Zain Mehdi}
 \email{zain.mehdi@anu.edu.au}
\affiliation{Department of Quantum Science and Technology and Department of Fundamental and Theoretical Physics, Research School of Physics, Australian National University, Canberra 2600, Australia}%
\author{Simon A. Haine}
\affiliation{Department of Quantum Science and Technology and Department of Fundamental and Theoretical Physics, Research School of Physics, Australian National University, Canberra 2600, Australia}%
\author{Joseph J. Hope}
\affiliation{Department of Quantum Science and Technology and Department of Fundamental and Theoretical Physics, Research School of Physics, Australian National University, Canberra 2600, Australia}%
\author{Stuart S. Szigeti}
\affiliation{Department of Quantum Science and Technology and Department of Fundamental and Theoretical Physics, Research School of Physics, Australian National University, Canberra 2600, Australia}%

\date{\today}

\begin{abstract}
We investigate the fundamental viability of cooling ultracold atomic gases with quantum feedback control.
Our study shows that the trade-off between the resolution and destructiveness of optical imaging techniques imposes constraints on the efficacy of feedback cooling, and that rapid rethermalization is necessary for cooling thermal gases. We construct a simple model to determine the limits to feedback cooling set by the visibility of density fluctuations, measurement-induced heating, and three-body atomic recombination. We demonstrate that feedback control can rapidly cool high-temperature thermal clouds in quasi-2D geometries to degenerate temperatures with minimal atom loss compared to traditional evaporation. Our analysis confirms the feasibility of feedback cooling ultracold atomic gases, providing a pathway to new regimes of cooling not achievable with current approaches.
\end{abstract}


\maketitle
The efficient cooling of quantum systems is a critical aspect of modern quantum science~\cite{Wieman1999,Leibfried2003,Aspelmeyer2014,Frowis2018}, including for precision measurement~\cite{Degen2017,Aspelmeyer2014}, quantum simulation~\cite{Bloch2008_Rev,Bauer2023}, and quantum information~\cite{Saffman2010,Brown2016}. A powerful tool for achieving this is feedback cooling, where the system is monitored and controlled in real-time~\cite{Zhang2017}. Although feedback cooling has been demonstrated for trapped ions~\cite{Bushev2006} and quantum mechanical oscillators~\cite{Cohadon1999,Guo2019,Rossi2018,Wilson2015,Li2011,Poggio2007,Magrini2021,Tebbenjohanns2021,Vovrosh2017}, its applicability to more complex, multi-mode systems, such as quantum gases, is still an open question. Rapid advances in the feedback control of ultracold atomic gases are expected, thanks to technological advances in highly-configurable optical potentials~\cite{Henderson2009,Gauthier2016,Gauthier2021}.
This presents an opportunity to explore the potential of feedback cooling for ultracold atomic gases, and to address the challenges and open questions related to this technique.

The potential benefits of feedback cooling ultracold atomic gases are significant. It may alleviate existing limits to achievable atom number ($\sim 10^7$ per shot) inherent to evaporative cooling~\cite{Wigley2016}, or improve current limits to achievable degeneracy for given total atom number. This latter aspect may play a critical role in space-based quantum sensing, where thermal expansion of the atomic cloud limits achievable sensitivity~\cite{Belenchia2022}. Limits to achievable degeneracy are even more pronounced for Fermi gases, with state-of-the-art demonstrations falling at least $10\times$ short of the degeneracy needed for Fermi-gas-based quantum simulators of high-temperature superconductivity~\cite{Onofrio2016}. \par 

However, a feedback-cooled ultracold atomic gas has not yet been experimentally realized, and it is unclear whether feedback cooling's efficacy is sufficient to achieve the above benefits. Although there have been theoretical investigations of feedback cooling in Bose gases~\cite{Haine2004,Wilson2007,Szigeti2009,Szigeti2010,Szigeti2013,Hush2013,Hurst2020,Wade2015,Wade2016,Schemmer2017}, the methods employed in these studies are computationally intensive, contextually bound, and usually limited to low-temperature gases in quasi-one-dimensional geometries, preventing studies into the viability of feedback cooling more generally. Furthermore, there have been no studies of feedback cooling ultracold Fermi gases, and as such the effect of quantum statistics on the feedback scheme is poorly understood.

Here we assess the viability of feedback cooling ultracold gases by establishing fundamental limits imposed by the physics of non-destructive optical imaging and the thermodynamics of thermal cold-atom ensembles.
We demonstrate the necessity of rapid rethermalization due to the constraints imposed by signal resolution and measurement destruction, concretely ruling out the possibility of feedback cooling single-component Fermi gases or one-dimensional integrable systems. We develop a continuous feedback-cooling model for thermal gases, including multi-component systems, and use this to show that large atom number ($N\sim 10^9$), highly-oblate atomic gases can be feedback cooled to degeneracy with minimal atom loss. Our results establish feedback as a promising number-conserving approach to cooling ultracold gases in reduced dimensions and sets a foundation for future investigations into feedback cooling single component Bose gases and Bose-Fermi mixtures.  \par

\begin{figure}[t!] \centering \includegraphics[width=\columnwidth]{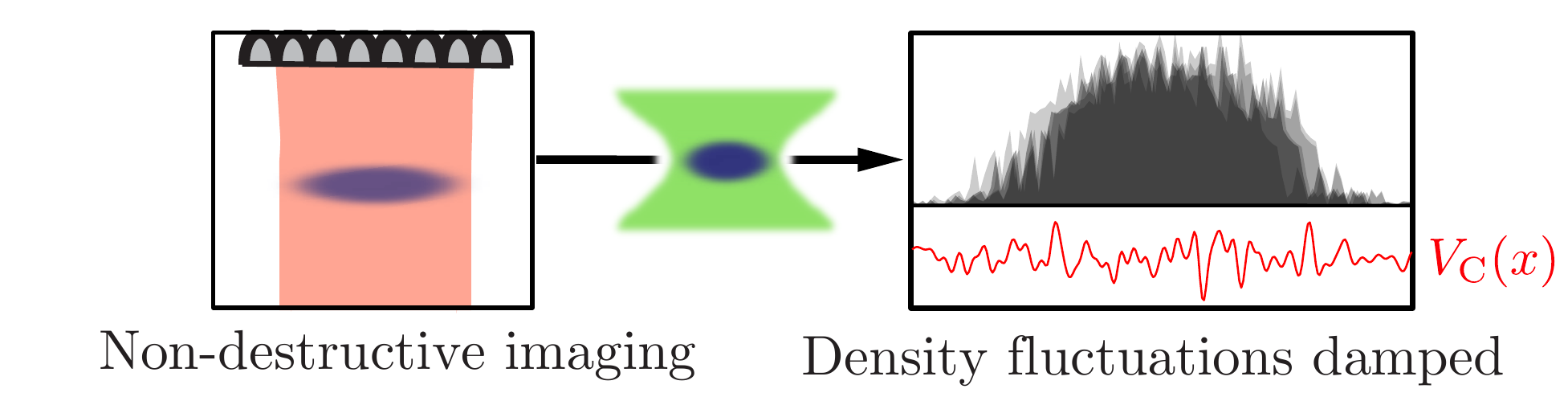}   \caption{Feedback loop. The atomic cloud density (dark blue) is non-destructively imaged by off-resonant light (red). Atomic density fluctuations ($1$D projection shown in grey) are then damped by a control potential $V_{\rm C}$ (red trace) realized, for e.g., by a configurable optical-dipole potential (green).}  \label{fig:FeedbackDiagram} \end{figure}

\SectionPRL{Feedback cooling scheme} We consider a realistic feedback cooling scheme (Fig.~\ref{fig:FeedbackDiagram}) based on models developed and validated in Refs.~\cite{Szigeti2009,Szigeti2010,Hush2013,Goh2022}. An atomic cloud is dispersively imaged by off-resonant coherent light illuminated along the tightly-trapped $z$-axis. This gives a real-time estimate of the atomic column density in the $xy$ plane, which is then used to construct a control potential that damps observable density fluctuations -- applied using a configurable optical dipole potential~\cite{Gauthier2016}, for example. For simplicity, we consider a cylindrically-symmetric geometry $\omega_x=\omega_y\equiv\omega_\perp$ with $\kappa\equiv \omega_z/\omega_\perp \gg 1$ and assume a Gaussian atomic density of radial and transverse standard deviations $R_\perp,\,R_z$, respectively.\par 
\emph{Direct control of thermal excitations.---} Without rethermalization, thermal excitations must be directly controlled, which is only possible if they are resolvable in the density image. Consequently, fundamental bounds on the measurement resolution give a parameter range for effective feedback cooling.

Firstly, the smallest possible lengthscale resolvable by optical imaging, $\Delta r$, is given by the gas `thickness' $R_z$ and light wavelength $\lambda$~\cite{Szigeti2009,Dalvit2002}:
\begin{align}
\label{eq:DiffractionLimit}
    (\Delta r)^2 \leq \frac{R_z \lambda}{2\pi} \equiv r_D^2\,.
\end{align}
This resolution limit arises since light rays cannot remain exactly parallel and diffract by a minimum angle as they pass through the gas. A second limit is set by the measurement strength, which must be sufficiently strong to accurately estimate the atomic density, yet sufficiently weak to avoid destruction of the atomic cloud via spontaneous emission. The signal-to-noise ratio (SNR) of a single pixel of area $(\Delta r)^2$ from any non-destructive optical measurement of the atomic density with classical light is bounded by the shot-noise limit \cite{Hope2004,Hope2005,Mehdi2024_Thesis}
\begin{align}
    \label{eq:SNR_CloseButNoHope}
    \text{SNR} \leq \bar{n}\sqrt{\eta (\Delta r)^2 \sigma_0\Gamma P_e \delta t } \,,
\end{align}
where $\eta\in (0,1]$ is the detector's quantum efficiency, $\sigma_0=3\lambda^2/(2\pi)$, $\Tilde{n}$ is the average column-density of the pixel, $\Gamma$ is the excited state linewidth, $P_e$ is the excited state probability, and $\delta t$ is the detector's integration time. The destruction associated with each measurement is quantified by the average number of spontaneous emission events per atom: $D\equiv \Gamma P_e \delta t$.

Useful feedback control requires $\text{SNR}\geq1$. By considering a perfect ($\eta = 1$) measurement of the atomic cloud with the minimum $\text{SNR}=1$ at a pixel with peak column density ($\overline{n} \leq \overline{n}_\text{max} = N / (2\pi R_\perp^2)$ for a Gaussian density), we can derive a lower bound on the amount of acceptable destruction. A conservative estimate of $\text{SNR}\geq1$ is given by taking the zero-temperature limits for the radii of ideal Fermi and Bose gases -- $R_{i} \approx\sqrt{\hbar/(m\omega_{i})} \left( 48 N \kappa \right)^{1/6}$~\cite{Giorgini2008} and $R_{i} =\sqrt{\hbar/(m\omega_{i})}$, respectively. Equation~\eqref{eq:SNR_CloseButNoHope} then implies
\begin{equation}
\label{eq:destruction}
    D  \geq \frac{4\pi^2\hbar^2}{(\Delta r)^2\sigma_0 m^2\omega_\perp^2 } 
\begin{cases}
    N^{-2},& \text{ bosons}\,,\\
   \left( 48 \kappa  \right)^{2/3}N^{-4/3}, & \text{ fermions} \,.
\end{cases}
\end{equation}
Thus, the required SNR for degenerate Fermi gases gives a factor
of $\left( 48 N \kappa \right)^{2/3}$ more destruction than for similar bosonic species, due to degeneracy pressure.  \par 

\begin{figure}[t!]
    \centering\includegraphics[width=\columnwidth]{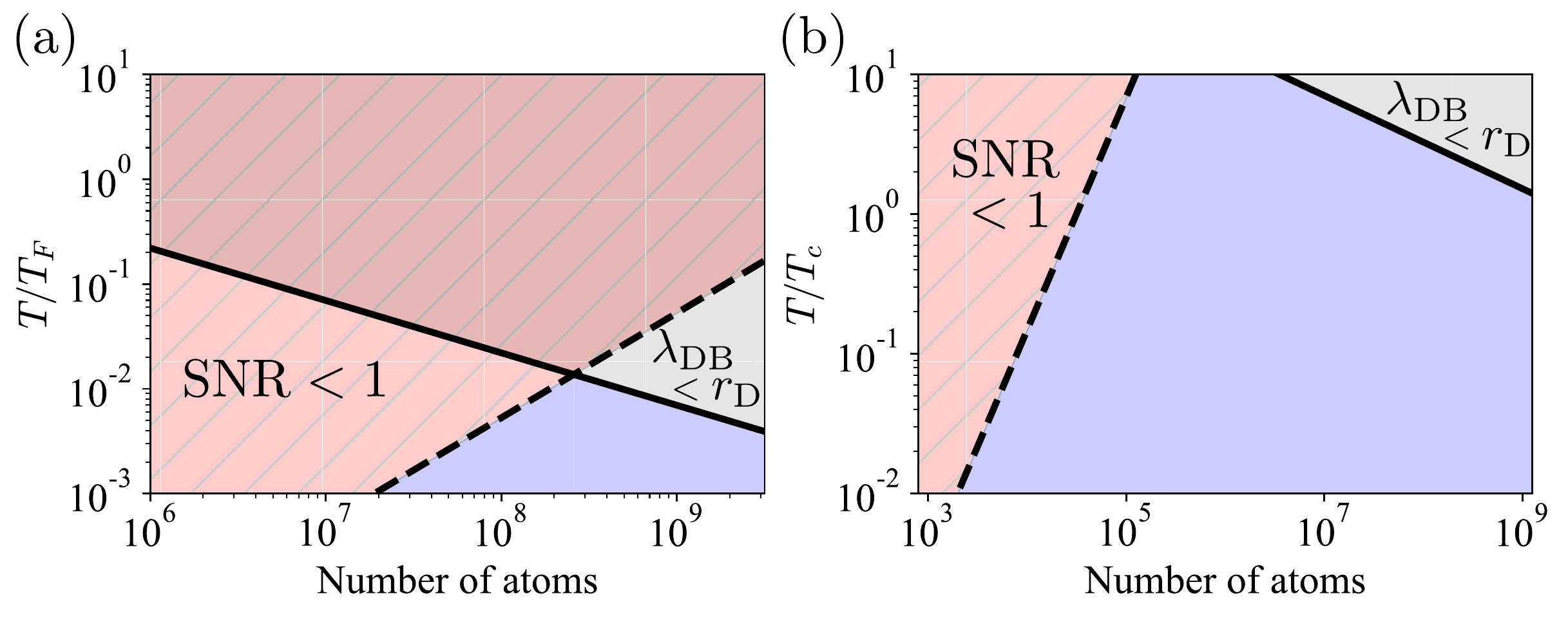}
    \caption{Regions where thermal excitations are resolvable for degenerate Fermi (a) and Bose (b) gases. Solid and dashed lines mark the temperature limits imposed by resolution (grey) and SNR (red), respectively. The $\text{SNR}\leq 1$ regime assumes a destruction limit of $D = 0.01$. The filled, blue area below both limits is where thermal density fluctuations are resolvable by the measurement. Temperature is in units of Fermi temperature $k_B T_F = \hbar \omega_\perp(6N\kappa)^{1/3}$ [(a)] and critical condensation temperature $k_BT_c^0\approx \hbar (N\kappa)^{1/3}\omega_\perp$ [(b)].    
     }
\label{fig:MaxTemp_Comparison_NonDim}
\end{figure}

Finally, we require a measurement resolution no larger than the thermal de~Broglie wavelength:  $\Delta r \leq \lambda_\text{DB}(T) = \sqrt{2\pi\hbar^2/(mk_B T)}$. Combining this with Eq.~\eqref{eq:DiffractionLimit} and Eq.~\eqref{eq:destruction} gives the regime where feedback cooling is effective (in the absence of rethermalization). To apply Eq.~\eqref{eq:SNR_CloseButNoHope}, we must choose an acceptable amount of destruction per measurement; we take$1\%$ absorption per measurement ($D=0.01$). For a concrete comparison, we consider $^6$Li (bosonic) and $^7$Li (fermionic) which have similar mass and electronic D1 ($\lambda=670$nm) transition~\footnote{Other commonly used atomic species have key parameters (mass and electronic transition wavelength) on the same order of magnitude}.

Figure~\ref{fig:MaxTemp_Comparison_NonDim} plots the regions bounded by our inequalities for trapping parameters $\{\omega_\perp/2\pi,\kappa\}=\{10~{\rm Hz},10^2\}$, and compares the degenerate Fermi and ideal Bose cases. The $(T,N)$ region where excitations are resolvable is much narrower for degenerate Fermi gases than degenerate ideal Bose gases. Since this narrow regime is beyond the capability of modern Fermi gas experiments, both in achievable atom number and temperature~\cite{Onofrio2016}, this rules out feedback cooling for single-component degenerate Fermi gases. In contrast, excitations are resolvable across the degenerate regime for Bose gases -- although, as shown below, this is not required for effective cooling due to rethermalization via interparticle scattering.

\SectionPRL{Thermodynamic model of feedback cooling}
Here we develop a thermodynamic model of feedback cooling, based on the premise that the controller can `only cool what it can see'. That is, the fraction of total energy removable depends on the excitations resolvable within the spatial resolution of the optical potential -- actuated by spatial light modulators (SLMs), for example. We first divide the atomic cloud into 2D cells of area $A$, each representing a single `controllable cell' of the configurable optical potential (e.g. a pixel on the SLM). Thus, $\sqrt{A}$ is the control's spatial-bandwidth. Each controllable cell contains $N_A = \bar{n} A$ particles, each contributing $3 k_BT$ of energy~\footnote{We neglect low-temperature corrections to the heat capacity of harmonically-trapped bosons, which are only significant near the critical condensation temperature $T\lesssim 2T_c$~\cite{Biswas2012}.}, giving total energy $E=3N_Ak_B T$ per cell. 

The spatial-bandwidth-limited control potential may only control the center-of-mass (COM) motion of each controllable cell in the $xy$ plane, which by equipartition contains energy $2k_B T$.
Thus, without rethermalization, feedback can at best extract a fraction of the total energy $\delta \equiv \Delta E/E \leq 2/(3N_A)$. $\Delta E$ is the energy removed by feedback; $\delta=1$ corresponds to removal of all energy from the cloud. This implies a single shot of feedback can extract the majority of the cloud's energy provided the area $A$ can be made arbitrarily small. However, the minimum value of $A$ is subject to two constraints. 
Firstly, $A$ is bounded from below by the spatial bandwidth of the SLM that actuates the controlling potential; $A \geq l_c^2$, where $l_c$ is the minimum achievable spatial resolution of the configurable optical potential. Secondly, effective cooling requires $A \geq (\Delta r)^2$, since excitations not resolvable by imaging cannot be controlled. Thus, $A=\max\{l_c^2,(\Delta r)^2\}$ is the optimal controllable cell size, which we choose for the remainder of this analysis.

The $A \geq (\Delta r)^2$ constraint can be combined with the imaging resolution limit, Eq.~\eqref{eq:DiffractionLimit}, to obtain $N_A \geq \bar{n} r_D^2$. For large atomic gases with $\bar{n}r_D^2 \gg 1$, this implies $\delta < 2/(3\bar{n} r_D^2) \ll 1$ -- i.e. a single shot of feedback can only remove a small fraction of the cloud's energy. Therefore, many iterations of feedback are required to cool the entire cloud; to accurately describe this, we must refine our estimate of $\delta$ to include imperfect control and measurement-induced heating.  

The achievable $\delta$ is limited by the finite measurement SNR, which misestimates the atom number in each pixel by $\Delta N_A=N_A/{\rm SNR}$. This prevents the perfect extraction of COM energy via feedback, giving residual COM energy $2(\Delta N_A/N_A) k_B T$ and thus a more realistic bound on the fraction of energy removable by a single iteration of feedback: $\delta\leq2(1-{\rm SNR}^{-1})/(3N_A)$. There are two fundamental channels of measurement-induced heating that further reduce the achievable $\delta$. Firstly, each spontaneous emission event heats the atomic cloud by twice the photon recoil energy, assuming a sufficiently deep trap such that the recoil does not induce loss. This contributes energy $2N_ADp_{\rm re}^2/m$ to each pixel per image, where $p_{\rm re}$ is the photon recoil momentum. Secondly, the fundamental measurement backaction heats the sample, with the atom-light interaction causing phase gradients in the atomic wavefunction that contribute kinetic energy (see Eq.~\eqref{eq:delta_backaction} in Appendix A). Together, these two heating channels increase the fraction of energy per pixel by $D\epsilon/(k_B T)$ with each measurement, where the energy-scale $\epsilon\equiv 4p_{\rm re}^2/(6m)+\pi\hbar^2/(4mR_z^2)$ includes the respective contributions from spontaneous emission and backaction. Since $R_z^2\gg \lambda^2$ in the thermal gas regime, spontaneous emission heating dominates over backaction (Appendix A), giving $\epsilon\approx 4k_B T_{\rm re}/3$, where $T_{\rm re}=p_{\rm re}^2/(2mk_B)$ is the recoil temperature.

Incorporating both measurement-induced heating contributions into the bound set by finite measurement SNR, we find that for $\mathcal{M}$ images taken per cycle of feedback, the best achievable net fraction of energy removed is:
\begin{align}
\label{eq:Delta_InclHeating} 	
\delta = \frac{2}{3 \bar{n}A}\left(1-{\rm SNR}^{-1}\right) - \mathcal{M}D\frac{\epsilon}{k_B T}.
 \end{align}
Since the upper bound on the SNR is proportional to $\sqrt{D}$, heating cannot be reduced arbitrarily by reducing $D$. Instead, we optimize the measurement strength -- parameterized by the destruction parameter, $D$ -- for each iteration of feedback. Assuming the SNR saturates Eq.~\eqref{eq:SNR_CloseButNoHope}, the optimal $D$ that maximizes Eq.~\eqref{eq:Delta_InclHeating} is:
\begin{align}
	D_{\rm opt} = \frac{1}{\bar{n} A}\left(\frac{A}{(\Delta r)^2}\frac{k_B^2 T^2}{9 \bar{n} \mathcal{M}^2\epsilon^2 \eta\sigma_0}\right)^{1/3} \,.
 \label{eq:Dopt}
\end{align}

Since cooling is only possible for $\delta \geq 0$, Eqs.~\eqref{eq:Delta_InclHeating} and \eqref{eq:Dopt} let us determine the minimum temperature achievable by feedback cooling (see Appendix B):
\begin{align}
T \geq T_{\rm min} \equiv \frac{A}{(\Delta r)^2}\frac{3\mathcal{M}}{8\eta \bar{n}\sigma_0} \frac{\epsilon}{k_B} \approx \frac{\mathcal{M}}{2\eta \bar{n}\sigma_0}T_{\rm re} \,,\label{eq:UltimateLimitT}
\end{align}
using $\epsilon\approx 4k_B T_{\rm re}/3$ and $A\approx (\Delta r)^2$~\footnote{$A\approx (\Delta r)^2$ can always be satisfied by fixing the imaging resolution to match the SLM spatial bandwidth: $\Delta r =  l_c$.}. This bound na\"ively suggests the temperature can be made arbitrarily small by making the column density $\bar{n}$ arbitrarily large, and increasing the detuning accordingly to keep the sample in the optically thin regime $\bar{n}\sigma_0 \ll 1+4\Delta^2/\Gamma^2$. However, $\bar{n}$ is limited by the peak 3D atom density via $n_{\rm 3D} = \bar{n}_{\max} /(\sqrt{2\pi}R_z)$, which cannot exceed a critical value $n_{\rm crit}$ set by three-body recombination losses \cite{Pethick2008}. As an example, consider a $^{87}$Rb cloud with $n_{\rm crit}\sim 10^{14}{\rm cm}^{-3}$; for $R_z\approx1\mu$m, Eq.~\eqref{eq:UltimateLimitT} gives $T_{\rm min}\approx 1$nK. This is well below typical transition temperatures ($\sim1\mu$K), suggesting feedback can cool thermal Bose gases to degeneracy. 

\SectionPRL{Continuous cooling with rethermalization}
Following feedback, the atomic cloud will rethermalize to a slightly lower temperature (provided $\delta >0$) $T_{\rm new}=T_{\rm old}(1-\delta)$~\footnote{This neglects the slight temperature dependence of the heat capacity near $T_c$.)} on a timescale set by the elastic collision rate $\Delta t^{-1} \equiv \gamma_{\rm el} = n_{\rm 3D}\sigma_s\sqrt{2}v_{\rm th}$, where $\sigma_s$ is the s-wave scattering cross-section and $v_{\rm th} = \sqrt{8k_B T/(\pi m)}$ is the thermal atomic velocity. For large atomic densities, each stage of feedback only reduces the energy by a small fraction $\delta \ll 1$. Therefore, for fast rethermalization we can take the infinitesimal limit of $(T_\text{new} - T_\text{old})/\Delta t$:
\begin{equation}
	\label{eq:ContinuousModeldTdt} \frac{dT}{dt} = -\gamma_{\rm el}(T)\delta (T) T \,, 
\end{equation}
where $\delta$ depends on $T$ through the implicit temperature dependence of the atomic density.

We bound the cooling rate $\gamma\equiv |dT/dt|/T$ using Eq.~\eqref{eq:ContinuousModeldTdt} and $\delta\leq 2/(3\bar{n}A)$:
\begin{align}
\label{eq:CoolingBound}
	\gamma \leq \frac{\sqrt{8}\sigma_s v_{\rm th}(T)}{3A }\frac{n_{\rm 3D}}{\bar{n}}\approx\frac{2\sigma_s v_{\rm th}(T)}{3\sqrt{\pi}A R_z} \,,
\end{align}
using $n_{\rm 3D}\approx\bar{n}/(\sqrt{2\pi }R_z)$ to relate the 3D atomic density to the measured column density, assuming a Gaussian profile. Then, since the control resolution is bounded by Eq.~\eqref{eq:DiffractionLimit}, we can find the upper bound to the cooling rate by substituting $A\geq r_D^2$:
\begin{align}
	\gamma \leq  \frac{4\sqrt{\pi}}{3}\frac{v_{\rm th}(T) \sigma_s }{ \lambda R_z^2} \,. \label{eq:UltimateSpeedLimit}
\end{align}

This expression is independent of atom number; while increasing the atom density via increasing $N$ speeds up rethermalization, it also reduces the fraction of energy stored in the COM motional energy of each pixel by the same amount (for fixed $A$). Equation~\eqref{eq:UltimateSpeedLimit} also demonstrates the importance of tight trapping in the imaging direction. Since we require $R_\perp \gg r_D$ for high-resolution spatial imaging, feedback cooling is most effective for highly oblate gases ($\kappa\gg 1$). For the above $^{87}$Rb example, $T=100\mu$K ($10\mu$K) gives $\gamma\lesssim 35$Hz ($110$Hz). 

\begin{figure}[t!]   \centering
   \includegraphics[width=.99\columnwidth]{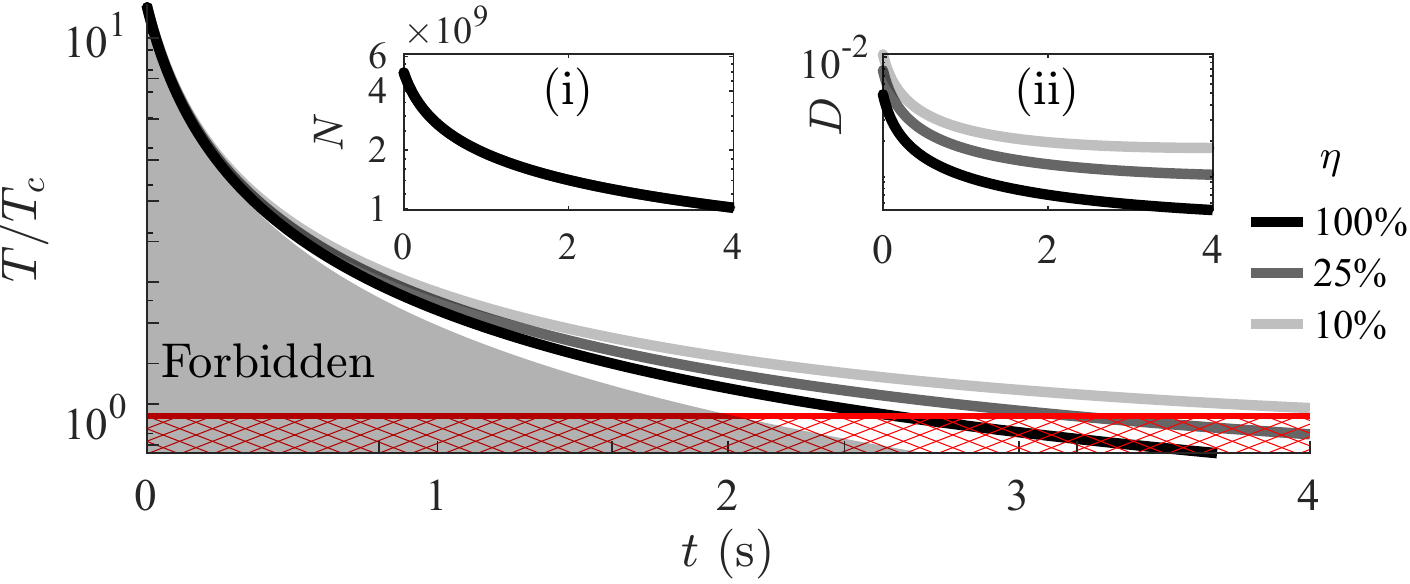}
    \caption{Feedback cooling an oblate $^{87}$Rb cloud of fixed spatial size $\{R_\perp,\kappa\} \approx \{597\mu{\rm m}, 180\}$ with initial number $N(0)=5\times10^9$ and temperature $T(0)=180\mu$K, for varied detection efficiencies $\eta$. Feedback can only extract energy visible in the control's spatial-bandwidth, here fixed at $l_c=1\mu$m; $\delta< 2/(3 N_A)$ defines a forbidden region (grey, shaded). (i) Three-body recombination causes $\sim 80\%$ of the initial sample to be lost over $4$s. (ii) Measurement destruction is optimized at each timestep to minimize heating. Calculations assume a thermal gas; below $T_c$ (red, hatched) the model breaks down.}    \label{fig:CoolingEx}
\end{figure}

\SectionPRL{Exemplary cooling demonstration} We have shown that feedback cooling is most effective when the atomic density is as large as three-body recombination permits, since this gives high SNR and fast rethermalization. High density can be maintained during cooling by reducing the trapping frequency as $\omega(t)=\omega(0)\sqrt{T(t)/T(0)}$, keeping $R_\perp,R_z$, and thus atomic density, constant. In Fig.~\ref{fig:CoolingEx} we use Eq.~\eqref{eq:ContinuousModeldTdt} to model such an experiment, where a $^{87}$Rb cloud with initial atom number $N(0)=5\times10^9$ and temperature $T=180\mu$K (typical parameters after Doppler cooling) is held in a trap with $\omega_\perp(0)/2\pi=35$Hz and fixed aspect ratio $\kappa=180$. 
The measurement strength (parameterized by $D$) is chosen at each time step using Eq.~\eqref{eq:Dopt} to optimize the trade-off between SNR and measurement-induced heating. We assume a spatial resolution of $l_c=1\mu$m for the control potential, well within the capability of SLMs based on digital-micromirror devices (DMDs), which have demonstrated sub-micron resolution~\cite{Weitenberg2011,Gauthier2016,Gauthier2021}. For these parameters, feedback cools the cloud to the critical temperature $T_c\sim 1\mu$K in $2-4$ seconds, while the atom number only reduces by $5 \times$ to $N \sim 10^9$ -- a performance impossible to achieve in alkali atoms with evaporative cooling.

Our feedback cooling protocol is robust to imperfect detector efficiency (or equivalently, technical imaging noise), with similar cooling achievable with $\eta=25\%$ (compared to perfect detection). For lower values of $\eta$, Fig.~\ref{fig:CoolingEx} shows the temperature plateauing at $\sim \mu K$ temperatures around $T_c$ due to the SNR approaching unity; cooling further would require a density increase that would give further atom loss.

\SectionPRL{Discussion} 
The temporal bandwidth of the feedback loop impacts feedback cooling if it is too slow, since density fluctuations cannot be cooled on timescales faster than the feedback loop's time delay. In such cases, it makes sense to increase the controllable pixel resolution $A$ to the size of the smallest density fluctuation resolvable within the feedback loop's temporal bandwidth (longer-wavelength excitations fluctuate on slower timescales). While this enables higher SNR imaging (Eq.~\eqref{eq:DiffractionLimit}), it reduces the rate of cooling by a factor of $A/l_c^2$, \emph{c.f.}~Eq.~\eqref{eq:CoolingBound}. We can determine whether this procedure is required by estimating the timescale of typical density fluctuations. The average time taken for atoms to move across a controllable cell is $\tau\equiv 2 \sqrt{A}/v_{\rm th}(T)\geq 2 l_c/v_{\rm th}(T)$. For $^{87}$Rb controlled at the fundamental imaging resolution limit (i.e. $A=r_D^2$), $T=10\mu$K ($200\mu$K) gives $\tau\approx 20\mu$s ($5\mu$s). This is within the temporal bandwidth of existing DMD-based SLMs -- e.g. Refs.~\cite{Gauthier2016,Gauthier2021} report switching speeds of $20$kHz.

Our model of continuous feedback cooling can be straightforwardly extended to describe the sympathetic cooling of multi-component atomic clouds via interaction with a feedback-cooled source. That is, the measurement and feedback cooling of a single component extracts energy from other components due to interatomic scattering between components and rethermalization. This could, for example, describe the feedback cooling of Bose-Fermi or Fermi-Fermi mixtures. For an $M$-component system of equal local density, each controllable cell contains $3MN_A$ degrees of freedom, yet only two COM degrees of freedom can be controlled. Following our earlier argument, the fraction of energy observable from optical imaging is a factor $M$ smaller than the single-component case (i.e. $\delta_{\rm multi}\leq 2/(3MN_A)$). Furthermore, $\sigma_s$ is reduced by $1/2$ for intercomponent scattering compared to scattering of identical particles, making the cooling rate in the sympathetic-cooling case $2M \times$ slower than if each component were measured and feedback-cooled individually (\emph{c.f.} Eq.~\eqref{eq:UltimateSpeedLimit}).

Our feedback model is constructed in the thermal gas regime; it breaks down qualitatively in the degenerate regime where quantum statistics affect rethermalization. For Bose gases, we expect feedback cooling to dramatically improve in the degenerate regime since the specific heat capacity sharply drops and there is Bose-enhanced transfer into the condensate mode. The increased role of interactions in degenerate Bose gases is also expected to improve the efficacy of feedback cooling, as coherence properties of the interacting atomic ensemble will be dominated by low-frequency, long-wavelength excitations~\cite{Kane1967} that are easier to control compared to the short-wavelength, high-frequency fluctuations of thermal gases. 
The opposite occurs for purely fermionic mixtures, as Pauli blocking reduces the elastic scattering rate~\cite{Holland2000,Crescimanno2000}. However, feedback could still prove advantageous by allowing much larger atom numbers than is accessible from evaporation. Furthermore, feedback could enhance sympathetic cooling of Fermi-Bose mixtures, which in the evaporative case is limited by the vanishing heat capacity of the Bose component~\cite{Onofrio2016}; extracting energy from the Bose gas via feedback could ameliorate this limitation and provide a pathway to achieving deeper Fermi degeneracy with large $N$. 

Our model's cooling can be realized by a multi-mode cold-damping control~\cite{Genes2008}, where optical control forces oppose the COM motion of the cloud at each `cell' of the spatiotemporal potential~\cite{Szigeti2010,Hush2013,Wade2016}. In the continuous limit, $l_c\rightarrow 0$, this control potential is $V_{\rm control}(x,t) \sim k \partial_t \Tilde{\rho}(x,t)$, where $\Tilde{\rho}(x,t)$ is the spatiotemporally-filtered measurement record and $k$ is the control gain -- referred to as the `energy-damping control' in Ref.~\cite{Goh2022}. In the degenerate regime, models of the partially-coherent atomic dynamics allow real-time state estimation and filtering~\cite{Szigeti2010,Szigeti2013,Hush2013,Dong2022}. This could enable more sophisticated control than permitted by our model, which may allow effective feedback with weaker measurement or better control with the same measurement SNR. 
Nevertheless, implementing a real-time filter of the full-field dynamics is a highly-challenging prospect; we expect near-term demonstrations will use simple spatiotemporal filters to improve the SNR of the real-time density estimates.

\SectionPRL{Conclusions} We have shown that feedback cooling could produce large atom-number quantum gases in low-dimensional geometries, beyond the fundamental capability of evaporative cooling. As our analysis is based on generic thermodynamic arguments and key operating principles of non-destructive optical measurement, these results are applicable to a broad range of experimental scenarios. Our work confirms that near-term proof-of-principle demonstrations are within the capability of established cold-atom experiments. 

We thank John Close for insightful discussions. This research was funded by the Australian Research Council project DP190101709. SAH acknowledges support through an Australian Research Council Future Fellowship Grant No. FT210100809. SSS was supported by an Australian Research Council Discovery Early Career Researcher Award (DECRA), Project No. DE200100495.

\section{Appendices}
\appendix
\section{A. Quantum backaction heating}
\renewcommand{\theequation}{A\arabic{equation}} \setcounter{equation}{0}
Here we estimate the heating effect due to fundamental quantum backaction. Our analysis is simplified by dividing the atomic density measurement into number measurements on spatial cells of area $r_D^2$, where $r_D$ is the fundamental resolution of the dispersive imaging, and treating each cell as being subjected to a quantum non-demolition (QND) number measurement. Measurement backaction scrambles the relative phase between cells, with a strength dependent on the number measurement precision. As a baseline, we expect the phase relation between unmeasured cells to be given by the magnitude of relative number fluctuations, $1/\sqrt{N_{\rm cell}}$. 

The QND measurement of a single cell is described by the unitary:
\begin{align}
	\hat{U}_{\rm QND} = \exp(i \beta \hat{N}_p \hat{N}_{\rm cell}) \,,
\end{align}
where $\hat{N}_p$ is the photon number operator and $\beta$ is a constant of proportionality we will relate to the shot-noise-limited SNR. Firstly, this unitary applies a phase to the light field $\phi_{\rm light} = \beta N_{\rm cell}$, with fluctuations given by the shot-noise relation~\cite{Taylor2016}:
\begin{align}
	\Delta \phi_{\rm light} = \frac{1}{2\sqrt{N_p}} = \beta \Delta N_{\rm cell} \,. \label{eq:shot-noise-relation}
\end{align}
Secondly, the QND unitary applies a phase $\phi_{\rm atom} = \beta N_p$ to the atomic state, with associated fluctuation $\Delta \phi_{\rm atom} = \beta \sqrt{N_p}$. Combining this with Eq.~\eqref{eq:shot-noise-relation} relates the magnitude of measurement-induced atomic phase fluctuations to the precision with which $N_{\rm cell}$ can be estimated:
\begin{align}
	\Delta \phi_{\rm atom} = \frac{1}{2\Delta N_{\rm cell}} =  \frac{\rm{SNR_{\rm cell} }}{2N_{\rm cell}} \,,
\end{align}
where $\rm{SNR_{\rm cell}}$ is the fundamental limit to the measurement sensitivity, given by the RHS of Eq.~\eqref{eq:SNR_CloseButNoHope} with $\Delta r=r_D$ and $\eta=1$. This expression demonstrates that stronger measurement gives larger phase fluctuations, and hence stronger backaction. For example, a perfect (projective) number measurement will give $\Delta N=0$ and hence infinitely strong phase fluctuations. 

To relate this to heating, we assume these phase fluctuations give rise to phase gradients $\partial_x \phi(x) \approx \Delta \phi_{\rm atom}/r_D$. This translates to a kinetic energy contribution, which we can compute by treating the gas as locally homogeneous over each cell. For a single cell this gives an energy contribution:
\begin{align}
	E_{\rm BA}^{\rm cell} &= -\frac{\hbar^2}{2m}\langle \int_{\rm cell} dxdy \hat{\psi}^\dag(\bm{x}) \nabla^2 \hat{\psi}(\bm{x})  \rangle  \\
	&\approx \frac{\hbar^2}{2m} N_{\rm cell}\int_{\rm cell} \frac{dxdy}{r_D^2} |\nabla \phi(\bm{x})|^2 \\ 
	&\approx \frac{\hbar^2}{m} N_{\rm cell} \left(\frac{\Delta \phi_{\rm atom}}{r_D} \right)^2 \,,
\end{align}
where $\hat{\psi}(\textbf{x})$ is the atomic field operator in the plane transverse to the imaging axis. The total energy added to each pixel is given by multiplying $E_{\rm BA}^{\rm cell}$ by the number of cells on each pixel $A/r_D^2$. Substituting in $N_{\rm cell} = N_A r_D^2 /A$,  $\Delta \phi_{\rm atom}   ={\rm SNR}/N_{\rm cell}$, and ${\rm SNR_{cell}}=\bar{n}\sqrt{ r_D^2 \sigma_0D}$, we can express this energy as:
\begin{align}
	E_{\rm BA}	&= \frac{A}{r_D^2} \frac{\hbar^2}{m} N_{\rm cell} \left(\frac{\rm SNR_{cell}}{N_{\rm cell} r_D} \right)^2 \\
&= N_A D\frac{\hbar^2}{2m} \frac{\sigma}{4 r_D^4}
\end{align}
where in the last line we have used $\bar{n}=N_A/A$. This takes the form of a free-particle dispersion relation with wavevector $\sqrt{\sigma_0}/(2r_D^2)$. Substituting the cross-sectional area of the light $\sigma_0=3\lambda^2/2\pi$ and resolution limit $r_D = \sqrt{R_z\lambda/(2\pi)}$, we see that the above relation depends only on the gas thickness:
\begin{align}
	E_{\rm BA} 
	&= 3 N_A \frac{\hbar^2}{4m} \frac{ D\pi}{R_z^2}
\end{align}
from which we obtain a heating contribution to $\delta$ of
\begin{align}
	\delta_{\rm BA} &= -\frac{\hbar^2}{4mR_z^2} \frac{\pi D}{k_B T} \,. \label{eq:delta_backaction}
\end{align}
We can compare this to the heating contribution from spontaneous emission $\delta_{\rm SE}$,  
\begin{align}
    \frac{\delta_{\rm BA}}{\delta_{\rm SE}} = \frac{3}{32\pi}\left(\frac{\lambda}{R_z}\right)^2 \,,
\end{align}
showing that spontaneous emission heating dominates over backaction heating if $R_z^2\gg\lambda^2$.

\section{B. Derivation of temperature bound}
\renewcommand{\theequation}{B\arabic{equation}} \setcounter{equation}{0}
Here we the temperature bound Eq.~\eqref{eq:UltimateLimitT}, in the main text. We start by substituting the optimal measurement strength -- parameterized by $D$, i.e. Eq.~\eqref{eq:Dopt} -- into Eq.~\eqref{eq:Delta_InclHeating} and solving for $\delta\geq 0$; the region where the increase in energy due to measurement-induced heating is greater than the amount of energy feedback can extract:\begin{align}
    \delta_{\rm opt} &= \frac{2}{3 \bar{n}A}\left(1-\frac{1}{\bar{n}\sqrt{\eta (\Delta r)^2 \sigma_0 D_{\rm opt} }}\right) - \mathcal{M}D_{\rm opt}\frac{\epsilon}{k_B T} \geq 0 \notag
    \,.
\end{align}
Solving the above inequality directly for $T$ does not lead to an insightful expression for the bound on the temperature. We may obtain a simple analytic expression by further assuming the column density $\bar{n}$ is sufficiently large that the shot-noise-limited SNR is significantly greater than unity, allowing the second bracketed term to be neglected. Then, solving the above inequality for $T$ yields Eq.~\eqref{eq:UltimateLimitT} of the main text:
\begin{align}
T \geq \frac{A}{(\Delta r)^2}\frac{3\mathcal{M}}{8\eta \bar{n}\sigma_0} \frac{\epsilon}{k_B} \,.
\end{align}

\section{C. Details of numerical simulation}
\renewcommand{\theequation}{C\arabic{equation}} \setcounter{equation}{0}
Here we outline the details of the numerical simulation used to produce Fig.~\ref{fig:CoolingEx}. The simulation is based on the continuous model Eq.~\eqref{eq:ContinuousModeldTdt} where the fraction of energy extracted per shot of feedback $\delta(T)$ is computed at each timestep using Eq.~\eqref{eq:Delta_InclHeating} with $\mathcal{M}=1$. These calculations use parameters corresponding to $^{87}$Rb (s-wave scattering length $a_s = 100.4a_0$, where $a_0$ is the Bohr radius), with the optical imaging far-detuned from the $D_2$ transition ($\lambda=780$nm). \par 

Our calculations also incorporate atom loss due to three-body recombination. We start with the rate equation describing three-body loss~\cite{Pethick2008}:
\begin{align}
	\frac{dn_{\rm 3d}}{dt} = -K n_{\rm 3d}^3 \,,
\end{align}
where $K$ is the three-body loss rate that is empirically determined. For $^{87}$Rb, $K\approx 4\times 10^{-41}$m$^6/$s~\cite{Everitt:2017} -- this sets the density scale at which loss becomes significant, $\mathcal{O}(10^{14})$cm$^{-3}$. We recast this equation in terms of the atom number $N$ using the Gaussian description of the atomic density employed in the main text, 
\begin{align}
\label{eq:dNdt_Loss} \frac{dN}{dt} &= - K N(t) \left(\frac{N(t)}{(2\pi)^{3/2}R_\perp^2R_z} \right)^2 \,.
\end{align}
We solve the coupled differential equations \eqref{eq:ContinuousModeldTdt} and \eqref{eq:dNdt_Loss} with a fourth-order Runge-Kutta algorithm (adaptive timestep). At each timestep, $\delta(T)$ is computed using Eq.~\eqref{eq:Delta_InclHeating} with $\mathcal{M}=1$ and $\bar{n}=N/(2\pi R_\perp^2)$.

\bibliography{bib}
\end{document}